# Energy Optimal Transmission Scheduling in Wireless Sensor Networks

Rahul Srivastava, *Student Member, IEEE,* and Can Emre Koksal, *Member, IEEE*

## Abstract

One of the main issues in the design of sensor networks is energy efficient communication of time-critical data. Energy wastage can be caused by failed packet transmission attempts at each node due to channel dynamics and interference. Therefore transmission control techniques that are unaware of the channel dynamics can lead to suboptimal channel use patterns. In this paper we propose a transmission controller that utilizes different "grades" of channel side information to schedule packet transmissions in an optimal way, while meeting a deadline constraint for all packets waiting in the transmission queue. The wireless channel is modeled as a finite-state Markov channel. We are specifically interested in the case where the transmitter has low-grade channel side information that can be obtained based solely on the ACK/NAK sequence for the previous transmissions. Our scheduler is readily implementable and it is based on the dynamic programming solution to the finite-horizon transmission control problem. We also calculate the information theoretic capacity of the finite state Markov channel with feedback containing different grades of channel side information including that, obtained through the ACK/NAK sequence. We illustrate that our scheduler achieves a given throughput at a power level that is fairly close to the fundamental limit achievable over the channel.

## I. Introduction

Energy efficient communication is one of the key concerns in the design of wireless sensor networks. Without any effort for adapting to the variability in the channel, the system resources are consumed inefficiently. For instance, if bad channel conditions are not anticipated, a high fraction of the node energy can be consumed by multiple retransmissions per correctly decoded packet. To avoid such inefficiencies, sensor nodes use transmission schedulers. The objective of an energy-efficient transmission scheduler is to reliably communicate data using minimal amount of energy, while meeting deadline and/or throughput constraints. Design of efficient transmission schedulers is challenging - especially in sensor networks - due to limited computational resources

The authors are with the Department of Electrical and Computer Engineering at The Ohio State University. Please direct all correspondence to Prof. Can Emre Koksal, Dept. ECE, 2015 Neil Ave., Columbus OH 43210, e-mail: koksal@ece.osu.edu, phone 614.688.4369. Rahul Srivastava can be reached at the same address/phone and e-mailed at srivastr@ece.osu.edu.

This work was supported by NSF grants 0635242 and 0831919.

and the lack of perfect channel side information (CSI) at the transmitter. Transmitters used in sensor networks often do not have access to advanced modulation techniques, which rules out the use of sophisticated adaptive transmission schemes.

Consequently we focus our attention on transmission schemes that do not require a high computational power. We propose a transmission scheduler that utilizes different "grades" of CSI on the state of a FSMC to schedule packet transmissions in order to meet a deadline constraint for the packets waiting in the transmission queue. We are specifically interested in the highly imperfect CSI that can be obtained based solely on the ACK/NAK sequence for the past transmissions. Note that, this level of information is available at the link layer in almost all wireless networks without any extra effort (such as sending special non-information carrying physical-layer pilot symbols over the channel). Our transmitter has a single transmit power level and the coding and modulation schemes are fixed. In every transmission opportunity, the transmitter can choose to attempt the packet transmission or defer it.

Our scheduler is based on the dynamic programming (DP) solution to a finite-horizon transmission control problem. We show how a simple version of it can be implemented in sensor nodes. We also calculate the capacity-cost function of the FSMC with feedback containing different grades of CSI. To our best knowledge, the capacity of channels with a feedback that is a random function of the channel state (e.g., ACK/NAK) has not been addressed before. We finally evaluate the performance of our scheduler and illustrate that it achieves a given throughput at a power level that is fairly close to the fundamental limit achievable over the channel.

In the transmission scheduling problem, the parameter we minimize is the number of transmission attempts, subject to a deadline constraint for the packets in the queue. One can realize that, with the limitation of binary power control (transmit or do not transmit), the number of transmissions is proportional to the energy consumed per correctly decoded packet and the deadline constraint can be translated into a throughput constraint at every point in time[1]. Hence the problem can be viewed as minimizing the energy per packet subject to a throughput constraint.

Transmission strategies based on channel estimation in FSMCs have been considered by Zorzi and Rao [1] as well as Chiasserini and Meo [2]. These papers assume a two-state Markov chain and detect the bad channel state upon receipt of a NAK. Both schemes reduce the transmission rate and constantly probe the channel as a response. In [2], the transmitter switches to a greedy

---

[1]For instance if the buffer contains 5 packets with a deadline constraint of 1 sec, then the scheduler has to guarantee a minimum of 5 packets/sec for the next second

transmission mode when the buffer level exceeds a certain threshold, regardless of the channel state. Our scheme is based on a dynamic program, which takes the queue and the estimated channel states jointly into account to schedule packet transmissions in an optimal manner. Johnston and Krishnamurthy [3] give an algorithm that minimizes the transmission energy and latency while transmitting over a fading channel by formulating the problem as a partially observed Markov decision process (POMDP) search problem. However, their threshold-based policy result is optimal for a 2-state channel only. In addition, the channel model is assumed to have a unity packet loss probability in one state. This may not seem like a fundamental difference, but in such a scenario, an ACK implies a "good" state. The corresponding time dependence is finite and the associated solution can exploit it. Our results are valid for a general FSMC.

Uysal, et. al., [4] and Zafer, et. al., [5] have considered rate control policies for transmission scheduling similar to ours. However, [4] uses a static channel model for deriving the control policy. On the other hand, [5] considers a Markov channel, but assumes knowledge of the channel state prior to transmission. In our model, this information is not available to the controller. Haleem, et. al., [6] use ACK/NAK feedback in a learning automata algorithm to schedule transmissions. This method is shown to converge to the optimal throughput in stationary channels. Ho, et. al., [7] give a sub-optimal rate adaptation scheme to maximize throughput that uses ACK/NAK feedback. Karmokar, et. al., [8] pose the problem of rate adaptation in Type-I Hybrid ARQ systems as a POMDP and provide some heuristic solutions. The authors develop this idea further and propose a linear programming approach to solve the POMDP problem [9], [10].

In the second part of the paper, we study fundamental limitations of FSMCs with feedback containing perfect and imperfect CSI. We will use tools from information theory to derive the capacity results for power constrained channels. Capacity of channels with CSI is a well studied topic. Goldsmith and Varaiya [11] gave the Shannon capacity of such channels with instantaneous CSI at the transmitter and receiver. In this case, CSI implies accurate knowledge of the instantaneous state of the Markov channel. The optimal power adaptation for such a channel model implies "water-filling" in time, which is analogous to the water-filling result in frequency in a frequency-selective channel [12]. Viswanathan [13] extended these results for delayed perfect CSI at the transmitter. Yüksel and Tatikonda derived the capacity of a FSMC with imperfect deterministic feedback [14]. Imperfect deterministic feedback implies that the transmitter has inaccurate (quantized) information about the channel state. The quantizer used is a deterministic function of the channel state. Although these results are useful for a wide variety of capacity calculations, they can not be applied directly to channels with discrete power levels and a long term average power constraint on the input symbols. This is due to the fact that the transmitter might have to defer transmission in order to meet the power constraints.

To incorporate skipped transmission attempts in the classical capacity results, we introduce a new discrete channel, which we call *channel with vacation*. We associate a channel with vacation to each state of the FSMC, and calculate the capacity-cost function for this special Markov channel. The general problem of finding the capacity of a feedback channel with memory has been notoriously difficult in information theory; and the problem of feedback capacity containing a CSI, which is a random function of the channel state has not been addressed before. We use this kind of a feedback in our channel model because the probability of a transmission success is channel state dependent, which implies that the feedback is a random function of the channel state. We generalize [14] and combine it with [13] and [15] in order to calculate the capacity-cost function of our channel with feedback. We compare this capacity-cost function with the performance achieved by our scheduler and observe that for a given throughput requirement, the power expended by our scheme is close to the theoretical limit of the system.

The paper is organized as follows. We give our system model and the assumptions in Section II. We present the DP formulation of the problem in Section III. We derive the theoretical bounds on the system performance in Section IV. In Section V, we analyze the performance of the DP scheduler and compare it with the theoretical bounds. We wrap up with a summary in Section VI.

## II. System Model

The system model used in this work is shown in Figure 1. Here, we consider a single point-to-point wireless link with a feedback path. The transmitter has $B$ packets in a queue to be sent to the receiver within $T$ time slots. The transmitter is assumed to transmit at a fixed rate and we define a time slot as the time it takes to transmit a packet and receive the associated ACK/NAK. The controller, which is attached to the transmitter, determines the decision to attempt a transmission or to defer transmission of a packet in every time slot. For the $k$th time slot, we denote this packet (or lack of it) as $x_k$. The receiver decodes the packet $y_k$ and sends feedback $c_k$ about the channel state to the controller over an error-free channel. The different types of feedback will be discussed shortly. At time $k$, the controller has access to the queue state information (i.e., number of packets in the queue, $q_k$) and the feedback $c_k$. We use the following notation for a sequence, $c_m^n \triangleq \{c_m, c_{m+1}, \ldots, c_n\}$. The information vector $I_k = \{q_0^k, c_0^k\}$ is defined as the information available at time $k$ to the controller. The controller makes control decision $u_k \in \{0, 1\}$ based on the information vector, $I_k$. Here '0' corresponds to the decision not to transmit and '1' corresponds to the decision to transmit packet $x_k$.

The channel model that we consider for this problem is a FSMC. The channel state at time $k$ is denoted by $s_k \in \mathcal{S}$, determines the packet loss probability during the $k$th time slot. The state space of the Markov channel, $\mathcal{S}$ is a discrete set containing the different channel states of the FSMC. Each channel state $s \in \mathcal{S}$ has a certain packet loss probability denoted by $\epsilon_s$. The channel state is assumed to be constant during the period of packet transmission, i.e., one time slot. The steady state probability of state $s$ is represented by $\pi(s)$. While not as simple as a Bernoulli or an independent loss channel model, Markov models are used to approximate channels with memory. FSMCs were first used to model a channel with bursty errors [16], and are popular in the literature [17]. We assume that the controller has accurate estimates of the channel parameters, i.e., transition probabilities and the packet loss probabilities associated with each state. Estimation of these channel parameters is beyond the scope of this paper. We note that this topic has been treated widely in literature, where an FSMC with unknown parameters is posed as a hidden Markov model (HMM). Iterative procedures for the Maximum Likelihood estimation of HMM parameters are well-understood, e.g., Baum-Welch algorithm [18].

The receiver is assumed to have an error detection scheme with a negligible probability of undetected error. The decoder identifies whether a packet is decoded correctly or not, i.e., $y_k = x_k$ or $y_k \neq x_k$. Packet error process at time $k$ is denoted by $z_k$, where $z_k = 1$ for a correct transmission and $z_k = 0$ for an incorrect transmission. The receiver sends the feedback containing the CSI to the controller on a channel assumed to be error-free. The three cases/grades of receiver feedback to the controller considered in this paper are:

1) **Non-causal Perfect CSI:** the controller knows the instantaneous channel state when it determines $u_k$, i.e., $c_k = s_k$ and $I_k = \{q_0^k, s_0^k\}$.

2) **Causal Perfect CSI:** the controller knows the delayed channel state when it determines $u_k$, i.e., $c_k = s_{k-1}$ and $I_k = \{q_0^k, s_0^{k-1}\}$.

3) **Causal Partial CSI:** in this case, the controller knows the ACK/NAK for the previous transmissions, i.e., $c_k = z_{k-1}$ and $I_k = \{q_0^k, z_0^{k-1}\}$.

For clarity of presentation, we use a unit delay for the causal CSI cases. The results contained in this paper can be easily extended to a general fixed delay. In the following section, we construct the finite-horizon DP solution. In Section IV, we keep the basic structure of this system model to derive the theoretical bounds.

## III. THE DYNAMIC PROGRAMMING SOLUTION

In this section we develop an algorithm/decision rule for the controller that minimizes the energy expended by the transmitter and the receiver for the successful delivery of a certain number of data packets while maintaining a deadline requirement. Intuitively, without a deadline constraint, the controller would be inclined to transmit at times only when it is almost sure of the good channel state to achieve energy efficiency. However, with the deadline constraint, if there are packets remaining in the queue close to the deadline, it may need to take chances with the bad state occasionally as well. Thus the controller has to make decisions on packet transmission, based jointly on the queue and channel states. Note that the controller needs to consider the effect of the current decision on future decisions. For instance, if the controller decides not to transmit at a time slot, there will be no feedback on the channel state for that time slot. This makes the available information more outdated, affecting the success of subsequent decisions made. In our system the CSI used by the controller is provided solely by the receiver. We assume that the initial queue state is also known by the receiver, hence the receiver also has the entire information vector. Consequently, the receiver has the knowledge of whether a transmission will be attempted at any given point in time and it can remain inactive to conserve energy during skipped transmission attempts.

Since the packet-loss process is a stochastic process, this problem can be viewed as one of sequential decision making under uncertainty. In our system model, both the queue evolution and channel evolution have a Markov structure. The queue state can be observed completely by the controller, however for the channel state, we have different cases of observation (complete and partial). The combination of these factors necessitates the use of a finite horizon DP approach [19] to achieve energy optimality and meet the deadline constraint at the same time.

### A. The Dynamic Program

Using the notation introduced in the previous section, we can write the state equation for the queue occupancy $q_k$ at the beginning of time slot $k = 0, 1, \ldots, T-1$, as follows,

$$q_{k+1} = f(q_k, s_k, u_k) = q_k - 1_{\epsilon_{s_k}} u_k, \tag{1}$$

where $u_k \in \{0, 1\}$. The packet loss process is denoted by $1_{\epsilon_{s_k}}$, where $1_{\epsilon_{s_k}} = 0$ with probability $\epsilon_{s_k}$ and $1_{\epsilon_{s_k}} = 1$ otherwise.

Our objective is to minimize the energy consumption of the transmitter while transmitting $B$ packets over $T$ time slots. Consequently, the cost function should be proportional to the total energy of the transmissions required to transmit all packets correctly. In order to do this, we set the cost incurred at time $k$, denoted by $g_k(q_k, s_k, u_k)$, equal to $u_k$. When the controller decides to transmit, it incurs a cost of 1 unit. On the other hand, if the controller decides to defer transmission, then it does not incur any cost. Since the packet loss is a stochastic process, there is always a non-zero probability associated with not transmitting all packets correctly by the deadline. To make this undesirable, we use a terminal cost $g_T(q_T) = Cq_T$, $C \gg 0$. Terminal cost implies that if there are $q_T$ packets left in the buffer at the end of time $T$, then the controller will incur a cost of $Cq_T$. One would expect that the decision rule and hence the performance is strongly tied to the choice of $C$, but we show that, somewhat surprisingly, in most simulations the choice of $C$ has little effect on the performance of the scheduler as long as $C \geq 10$. We can express the expected cost incurred from the $(T-1)$th stage to termination, also called the *cost-to-go function* $J_{T-1}(I_{T-1})$ as,

$$J_{T-1}(I_{T-1}) = \min_{u_{T-1} \in \mathcal{U}} \left\{ \mathop{\mathrm{E}}_{s_{T-1}} \left[ g_T(f(q_{T-1}, s_{T-1}, u_{T-1})) + g_{T-1}(q_{T-1}, s_{T-1}, u_{T-1}) | I_{T-1}, u_{T-1} \right] \right\}$$

$$= \min_{u_{T-1} \in \mathcal{U}} \left\{ u_{T-1} + \mathop{\mathrm{E}}_{s_{T-1}} \left[ Cf(q_{T-1}, s_{T-1}, u_{T-1}) | I_{T-1}, u_{T-1} \right] \right\}. \tag{2}$$

The cost-to-go $J_k(I_k)$ for stages $k = 0, 1, \ldots, T-2$, can be expressed iteratively by using successive cost-to-go functions $J_{k+1}(I_{k+1})$,

$$J_k(I_k) = \min_{u_k \in \mathcal{U}} \left\{ \mathop{\mathrm{E}}_{s_k, c_{k+1}} \left[ J_{k+1}(I_k, q_{k+1}, c_{k+1}, u_k) + g_k(q_k, s_k, u_k) | I_k, u_k \right] \right\}$$

$$= \min_{u_k \in \mathcal{U}} \left\{ u_k + \mathop{\mathrm{E}}_{s_k, c_{k+1}} \left[ J_{k+1}(I_k, q_{k+1}, c_{k+1}, u_k) | I_k, u_k \right] \right\}. \tag{3}$$

We note that these optimization problems reduce to a DP with perfect state information when non-causal perfect CSI is available. In case of causal CSI (perfect and partial) these problems are treated as DP with imperfect state information. At each $k = 0, 1, \ldots, T-1$, the optimal policy $\mu_k^*(I_k)$ maps the information vector $I_k$ to the control action $u_k \in \{0, 1\}$ that minimizes the cost-to-go given in (2) and (3). The optimal policy is obtained recursively, we first solve the optimization problem (2) for all possible values of the information vector $I_{T-1}$ to get $\mu_{T-1}^*(I_{T-1})$. The corresponding value of $J_{T-1}^*(I_{T-1})$ is used to calculate $\mu_{T-2}^*(I_{T-2})$ in (3) and $J_{T-2}^*(I_{T-2})$. This process is continued till $J_0^*(I_0) = J_0^*(q_0, c_0)$ is found. We note that, in general, at each stage $k$ the state space size of $I_k$ will grow exponentially. However, if the components of $I_k$ follow a Markov transition (e.g., perfect CSI feedback) then only the most recent state observations are required for determining the control decision keeping the state space size constant.

Next, we derive the DP equations for the case when causal partial CSI is available to the controller. The DP equations for causal and non-causal perfect CSI are simple applications of (2) and (3) and are omitted due to space constraints.

## B. DP Equation for Causal Partial CSI at the Controller

Recall that with causal partial CSI, $I_k = \{q_1^k, z_1^{k-1}\}$. The queue state, $q_k$, follows a Markov transition, so we can reduce the information vector to $I_k = \{q_k, z_1^{k-1}\}$. We cannot apply the same reduction to the ACK/NAK sequence, $z_k$. This is due to the fact that, even though $z_k$ is a function of the channel state $s_k$, in general, the probability distribution of $z_k$ does not depend only on $z_{k-1}$ but all the past observations, $z_1^{k-1}$. The problem with directly applying the DP approach, (2) and (3), in this case is that the state space of the information vector will expand exponentially with the received observations. To avoid this problem, we introduce a new quantity $\boldsymbol{w_k} = \{w_k(1), \ldots, w_k(|\mathcal{S}|)\}$ which is the conditional state distribution given the past sequence, i.e., $w_k(s) \triangleq p\left(s_k = s \mid z_1^{k-1}\right)$. The conditional state distribution follows a Markov transition, $\boldsymbol{w_k} = \Phi(\boldsymbol{w_{k-1}}, z_{k-1}, u_{k-1})$, which can be derived using a straightforward application of Bayes' rule. Since this quantity depends on observations and control actions from the previous stage only, the controller needs to track only the most recent value of this variable. Evaluating the expectation in (2) and (3), using $\boldsymbol{w_k}$, we write the cost-to-go function $J_k$. For the terminal stage,

$$J_{T-1}(q_{T-1}, \boldsymbol{w_{T-1}}) = \min_{u_{T-1}} \left\{ u_{T-1} + \sum_{s \in \mathcal{S}} w_{T-1}(s) \left( \epsilon_s C q_{T-1} + (1 - \epsilon_s) C(q_{T-1} - u_{T-1}) \right) \right\}, \quad (4)$$

and for the intermediate stages, $k = 0, \ldots, T-2$,

$$J_k(q_k, \boldsymbol{w_k}) = \min_{u_k} \left\{ u_k + \sum_{s \in \mathcal{S}} w_k(s) \epsilon_s J_{k+1}(q_k, \Phi(\boldsymbol{w_k}, 0, u_k)) \right.$$
$$\left. + \sum_{s \in \mathcal{S}} w_k(s)(1 - \epsilon_s) J_{k+1}(q_k - u_k, \Phi(\boldsymbol{w_k}, 1, u_k)) \right\}. \quad (5)$$

Note that (4) and (5) are valid for any discrete set $\mathcal{S}$. The recursive relation for $\boldsymbol{w_k}$ is given by,

$$\boldsymbol{w_k} = \Phi(\boldsymbol{w_{k-1}}, z_{k-1}, u_{k-1}) = \frac{\boldsymbol{w_{k-1}} \mathrm{A}(z_{k-1}, u_{k-1}) \mathrm{P}}{\boldsymbol{w_{k-1}} \mathrm{A}(z_{k-1}, u_{k-1}) \mathbf{1}^T}, \quad (6)$$

with $\boldsymbol{w_0}$ initialized as the steady state distribution of states, $\{\pi_s, \forall s \in \mathcal{S}\}$. Here $\mathrm{A}(z_{k-1}, u_{k-1})$ is a diagonal $|\mathcal{S}| \times |\mathcal{S}|$ matrix with $i$th diagonal term defined as,

$$p\left(z_{k-1} \mid S_{k-1} = i, u_{k-1}\right) = \begin{cases} \epsilon_i, & z_{k-1} = 0, u_{k-1} = 1 \\ 1 - \epsilon_i, & z_{k-1} = 1, u_{k-1} = 1 \\ 1, & u_{k-1} = 0 \end{cases},$$

P is the transition probability matrix for the FSMC, and $\mathbf{1} = \{1, \ldots, 1\}$ is a $|\mathcal{S}|$-dimensional vector. We note here that when $u_{k-1} = 0$, there is no transmission in the $(k-1)$th slot and $z_{k-1}$ is not known. When this "gap" appears in the ACK/NAK sequence, $\boldsymbol{w_k}$ is updated by vector multiplication with the transition matrix P. Intuitively, the conditional state estimate will be less accurate when there are gaps in the ACK/NAK sequence which leads to a degradation in the performance of the DP-scheduler. As expected, with increasing gap size, the state estimate converges to the steady state probability.

*C. Implementation Issues*

In this section, we discuss some implementation issues associated with the DP-scheduler. The computation required to solve the DP equations is beyond the computational capability of most sensor nodes. To bypass this problem, we propose the use of lookup tables to implement the scheduler. This approach requires memory resources rather than computation resources. A lookup table $\mathrm{T}_{\mathcal{U}}(\mathrm{P}, \epsilon_1, \epsilon_2, \ldots, \epsilon_{|\mathcal{S}|}, k, q_k, \boldsymbol{w_k})$ which has been pre-computed and loaded in the sensor memory is used to determine the control rule. The table contains the control action $u_k$ for different channel and queue conditions. The controller updates $\boldsymbol{w_k}$ according to Eq. (6) and looks up the control action, $u_k$, corresponding to this vector. Since it is not practical to have such a table for all channel realizations, we store tables for a discrete set of channel realizations, $\mathcal{A}$.

The major design parameters on the transmitter/controller in this approach is the amount of memory to store the look-up table $\mathrm{T}_{\mathcal{U}}(\mathrm{P}, \epsilon_1, \epsilon_2, \ldots, \epsilon_{|\mathcal{S}|}, k, q_k, \boldsymbol{w_k})$, where $\{\mathrm{P}, \epsilon_1, \epsilon_2, \ldots, \epsilon_{|\mathcal{S}|}\} \in \mathcal{A}$. The size of memory will depend on the number of channel realizations $|\mathcal{A}|$ stored in the look-up table. In addition, each value of $q_k$ and $k$ will have a set of $\boldsymbol{w_k}$ associated with it. Let each component of $\boldsymbol{w_k}$ be quantized into 10 levels. If the DP algorithm provides a solution for up to 100 data packets transmitted over a duration of up to 500 time units, the memory requirement will be $|\mathcal{A}| \times 50 \times 10^{|\mathcal{S}|-1}$ Kbits (since each the control action will need only one bit of storage). For $|\mathcal{A}| = 100$ and $|\mathcal{S}| = 3$, the memory requirement for the controller will be 62.5 Mbytes.

Before concluding this section, we would like to point out that there is an inherent tradeoff between accuracy and complexity with increasing number of states. The memory requirements for storing the controller action will increase exponentially with the number of FSMC states. On the other hand, the channel model will become more "accurate" as more channel states are taken into account. However, there is no universal rule for selecting the number of states in the FSMC model for all possible physical fading processes. We direct the reader to [17] and references therein for a discussion on the selection of number of FSMC states.

## IV. Performance Bounds

The DP-based scheduler in Section III provides a strategy to consume the minimum energy, $\mathcal{P}T$, to attempt successful transmission of $B$ packets by a specified deadline $T$. This corresponds to a throughput of $B/T$ at power $\mathcal{P}$. In this section, we relax the deadline constraint and find the maximum achievable *asymptotic* throughput subject to a power constraint $\mathcal{P}$. This quantity is known as the capacity-cost function $C(\mathcal{P})$ [20]. The inverse of the capacity-cost function gives the minimum power $\mathcal{P}$ for a given asymptotic throughput $C$ [21], linking it to the objective of the DP problem. Inverting the axes of $\mathcal{P}$-$C(\mathcal{P})$ plot will give the plot of $\mathcal{P}$ as a function of $C$.

The capacity-cost function will give an upper bound on the achievable throughput of the DP-scheduler. To understand this statement, we note that the information theoretic calculations are asymptotic results, i.e., number of channel uses tend to infinity while meeting a throughput requirement. Any scheduler that transmits $B$ packets in $T$ time slots will achieve an asymptotic throughput of $B/T$ packets/time slot by replicating itself every $T$ time slots. However, the converse is not true, as ensuring an asymptotic throughput does not guarantee a throughput for every block of $T$ time slots. As a result, the capacity-cost function requires less power to achieve the same throughput and provides an upper bound on the throughput of the DP-scheduler.

To adapt our system for information theoretic calculations, we make some modifications to the system model shown in Figure 2. Since there is no deadline constraint in the problem, the queue in this system is assumed to be infinitely backlogged, and omitted from the figure. The channel state, as before, is a Markov chain $\{S_k \in \mathcal{S}, \ k = 1, 2, \ldots\}$ and the CSI $C_k$ is a function of $S_k$. The channel input is denoted by $X_k \in \mathcal{X}$ and output is denoted by $Y_k \in \mathcal{Y}$. The messages, chosen from a set of equiprobable set of messages $\mathcal{W}$, are denoted by $W$ and the decoded messages are denoted by $\hat{W}$. The size of the message set $\mathcal{W}$ is $M$. Before presenting the specifics of the capacity calculation, we define some information theoretic quantities.

*Encoder:* For a given message set $\mathcal{W}$, an encoder is a sequence of code functions $\{f_k(w, c_0^{k-1})\}_{k \geq 1}$, where the symbol to be sent at time $k$ is given by $f_k(w, c_0^{k-1})$.

*Channel Code:* For a given message set $\mathcal{W}$, the $(T, M, \mathcal{P}, e)$ channel code consists of: **(1)** The blocklength of the codewords is equal to $T$, i.e., the number of channel uses to transmit a message is equal to $T$. **(2)** Each codeword $\{f_k(w, c_0^{k-1})\}_{1 \leq k \leq T} \ \forall \ w, \ c_0^{T-1}$, satisfies the constraint, $\frac{1}{T} \sum_{i=1}^{T} b(f_i(w, c_0^{i-1})) \leq \mathcal{P}$, where $b(x)$ is the non-negative cost associated with the input symbol $x \in \mathcal{X}$. Therefore $\mathcal{P}$ is a hard power constraint imposed on every sample outcome of the channel state or ACK/NAK sequence. **(3)** The average probability of incorrectly decoding a message is bounded as $\frac{1}{|\mathcal{W}|} \sum_{w \in \mathcal{W}} p(\hat{w} \neq w | W = w) \leq e$.

*Capacity-Cost Function:* Given $0 \leq e < 1$ and $\mathcal{P} > 0$, a non-negative number $R$ is an $e$-achievable rate with the average cost per symbol not exceeding $\mathcal{P}$ if for every $\delta > 0$ there exists $T_0$ such that if $T \geq T_0$, then an $(T, M, \mathcal{P}, e)$ code can be found whose rate satisfies $\log M > T(R - \delta)$. Furthermore, $R$ is said to be achievable if it is $e$-achievable for all $0 < e < 1$. The maximum achievable rate with average cost per symbol not exceeding $\mathcal{P}$ is the channel capacity denoted by $C(\mathcal{P})$, which is referred to as the capacity-cost function.

We cannot directly apply classical channel capacity results as the system that we consider has special properties. The factors to consider while evaluating the channel capacity are:

1) **Skipped Channel Uses:** An ON-OFF scheduler is limited to two discrete power levels corresponding to the decision to transmit (at a fixed power) or not to transmit. As a result, under a power constraint, we should consider an "outage capacity" in which the decision not to transmit is accounted for in the capacity calculations.

2) **Markov Channel with Delayed Feedback:** Deterministic perfect and quantized feedback in such channels has been treated in literature [13], [14], but randomly quantized feedback has not. We have to calculate the capacity-cost function of a Markov channel with feedback in the form of ACK/NAK. In our channel model, ACK/NAKs are a random function of the channel state and hence need to be considered accordingly.

### A. Finite State Markov Erasure Channel (FSMEC)

A FSMEC is a special case of FSMC where, during each symbol duration is one of a finite number of erasure channels (ECs) with Markov transitions between these ECs. The channel state $S_k$ determines the channel erasure probability during the $k$th symbol. Note that we describe an entire packet (or lack of it) as a channel symbol and we model a detectable packet error using an erasure. In this problem, since we assume an ON-OFF transmitter, the input symbols are restricted to discrete power levels and have a long term average power constraint. Consequently, it is not possible to transmit data at every time instant. Discrete ECs [12] do not provide any provision for calculating mutual information rate under such power constraints. As a result, the channel model has to be designed in such a way that it incorporates the "OFF-period" of the transmitter in the capacity calculations.

To integrate the idea of no transmission in our channel model, we add an input symbol ('$\mathcal{V}$') to the traditional EC as shown in Figure 3 and call the new channel erasure channel with vacations (ECV). Each input symbol has an associated cost (power) of 1 unit and the cost required to transmit '$\mathcal{V}$' is 0. The transition probabilities of '$\mathcal{V}$' are chosen according to the statistical requirement rather than the physical signaling system. We will show in Appendix A that the capacity-cost function $C(\mathcal{P})$ of the ECV in Figure 3 is $\mathcal{P}N(1 - \epsilon_s)$. This can be intepreted as the capacity of an EC that is active for $\mathcal{P}$ fraction of time. One can observe that the unconstrained capacity (i.e., $\mathcal{P} = 1$) of this ECV is the same as that of an EC with $2^N$ input symbols and erasure probability $\epsilon_s$. As a result, addition of '$\mathcal{V}$' to the EC does not change the mutual information between the input and output symbols. Therefore, we infer that '$\mathcal{V}$' is a zero-cost, zero-information symbol which can replace the OFF-period of the transmitter for the capacity calculation. For the ON-period, we consider transmission of $N$-bit (fixed-size) packets that require unit cost to transmit. Hence, there will be $2^N$ distinct symbols, $\{`1`, \ldots, `2^N`\}$, shown in Figure 3 representing the ON-period of the transmitter. To make this model equivalent to the one in Section II, we make the transition and the error probabilities of the two channels equal. We assume that the probability of packet error is independent of the packet "content". As a result in state $s$, each packet (symbol) is transmitted correctly with probability $1 - \epsilon_s$.

### B. Capacity of the FSMEC

We state the following results for the capacity-cost function of the FSMEC with causal perfect and partial CSI at the transmitter. Each state $s \in \mathcal{S}$ is associated with an ECV (shown in Figure 3) with parameter $\epsilon_s$, steady state probability $\pi(s)$ and $2^N + 1$ input symbols.

*Theorem 1:* The capacity-cost function for a FSMEC with transition probability matrix P, channel state space $\mathcal{S}$, and causal perfect CSI at the encoder is given by,

$$C_{\text{FSMEC-CSI}}(\mathcal{P}) = \sup_{\mathcal{P}(\tilde{s})} \sum_{\tilde{s} \in \mathcal{S}} \sum_{s \in \mathcal{S}} \pi(\tilde{s}) \text{P}(\tilde{s}, s)(1 - \epsilon_s) N \mathcal{P}(\tilde{s}),$$

$$\text{s.t.} \quad \sum_{\tilde{s}} \pi(\tilde{s}) \mathcal{P}(\tilde{s}) \leq \mathcal{P} \quad \& \quad 0 \leq \mathcal{P}(\tilde{s}) \leq 1, \quad \forall \tilde{s} \in \mathcal{S}. \tag{7}$$

The optimization variable is $\mathcal{P}(\tilde{s})$, which can be thought of as the power allocation policy for state $\tilde{s}$. It is the fraction of unit cost symbols in the codebook for state $\tilde{s}$.

The solution to (7) gives the optimal power allocation policy $\{\mathcal{P}^*(s), \forall s \in \mathcal{S}\}$. Substituting this power allocation into the original expression gives the channel capacity. Following is a brief sketch of the proof, details of which can be found in Appendix A. Each ECV, associated to a state in the FSMEC, has a capacity-cost function. First we find the capacity-cost function of the ECVs using the expression given by Verdu [15]. Next, we use the capacity expression for a FSMC with causal perfect CSI at the transmitter given by Viswanathan [13]. The capacity-cost function substituted in the FSMC capacity expression leads to the optimization problem (7).

We present similar results when causal partial CSI is available to the encoder. We denote $z_1^n \in \{0,1\}^n$ as an ACK/NAK sequence of length $n$ and $p(z_1^n)$ is the probability of $Z_1^n = z_1^n$.

*Corollary 2:* The capacity-cost function for FSMEC with each state $s \in \mathcal{S}$ associated with an ECV with parameter $\epsilon_s$ and causal partial CSI (ACK/NAK) at the encoder is given by,

$$C_{\text{FSMEC-ARQ}}(\mathcal{P}) = \lim_{n \to \infty} C^n(\mathcal{P}) = \lim_{n \to \infty} \left[ \sum_{\substack{z_1^n \in \{0,1\}^n \\ s_{n+1} \in \mathcal{S}}} p(z_1^n) \sup_{\mathcal{P}(z_1^n)} \left\{ p(s_{n+1}|z_1^n)(1-\epsilon_{s_{n+1}}) N \mathcal{P}(z_1^n) \right\} \right]$$

$$\text{s.t.} \quad \sum_{z_1^n} p(z_1^n)\mathcal{P}(z_1^n) \leq \mathcal{P} \quad \& \quad 0 \leq \mathcal{P}(z_1^n) \leq 1, \quad \forall z_1^n \in \{0,1\}^n. \tag{8}$$

Here, the optimization is done over $\mathcal{P}(z_1^n)$, which can be thought of as the power allocation policy for the ACK/NAK observation $z_1^n$.

This corollary extends Theorem 1 to the case when only ACK/NAK information is available to the encoder. In this case the power allocation policy becomes a function of the observed ACK/NAK sequence instead of the last observed state. The first part of the proof, i.e., capacity-cost calculation, remains unchanged. We change the feedback capacity expression used in Theorem 1 to one adapted from [14]. The proof is mostly kept intact, a modification is made in the formulation of Fano's inequality in Eq. (12) of [14] to incorporate ACK/NAK feedback at the transmitter. The details of the modification are given in Appendix B.

The particular expression (8) provides an indication to a level-filling solution to the optimization problem. To evaluate this expression, we calculate the joint probability $p(s_{n+1}, z_1^n)$. The joint probability expression can be expanded as $\sum_{s_n} p(s_{n+1}|s_n)p(z_n|s_n) \cdots \sum_{s_1} p(s_2|s_1)p(z_1|s_1)p(s_1)$, which can be evaluated iteratively. The optimization problem is then solved over all possible sequences $z_1^n$ which results in a level-filling solution for $2^n$ states. We calculate $C^n(\mathcal{P})$ for different values of $n$ and observe that, for most cases, $C^{10}(\mathcal{P})$ is reasonably close to $C_{\text{FSMEC-ARQ}}(\mathcal{P})$.

*Example 1:* (A 2-state Markov erasure channel (2SMEC) with perfect causal CSI) We consider a 2SMEC with $\epsilon_1 < \epsilon_2$ and transition probabilities $\text{P}(1,2) = p_{12}$ and $\text{P}(2,1) = p_{21}$. For this channel, $\pi_1 = \frac{p_{21}}{p_{12}+p_{21}}$ and $\pi_2 = \frac{p_{12}}{p_{12}+p_{21}}$. Applying Theorem 1, we can find the power allocation policy $\{\mathcal{P}(1), \mathcal{P}(2)\}$. The plot for $\{\mathcal{P}(1), \mathcal{P}(2)\}$ as a function of $\mathcal{P}$ is shown in Figure 4. For Case 1, when $(1-p_{12})(1-\epsilon_1) + p_{12}(1-\epsilon_2) \geq p_{21}(1-\epsilon_1) + (1-p_{21})(1-\epsilon_2)$ and power constraint $\mathcal{P} \leq \pi_1$, power is allocated only to $\tilde{s} = 1$. This is represented by the solid blue trace. For $\mathcal{P} > \pi_1$, there are more opportunities to transmit then there are "good states" available. As a result, $(\mathcal{P} - \pi_1)$ part of the power is now allocated to $\tilde{s} = 2$ shown by the dashed red trace. This

 

continues till $\mathcal{P} = 1$, at which point all the states are utilized for transmission. A corresponding trend can be shown when $(1 - p_{12})(1 - \epsilon_1) + p_{12}(1 - \epsilon_2) < p_{21}(1 - \epsilon_1) + (1 - p_{21})(1 - \epsilon_2)$. These traces can be interpreted as level-filling results for power allocation in discrete channels, analogous to the one in [11].

## V. Performance Evaluation

### A. Scheduler Performance and Theoretical Bounds

We conduct simulations on FSMCs to compare the various transmission schedulers and compare them against their respective theoretical bounds. The metric used to compare the scheduler performance is termed *normalized transmission cost* (NTC). We define $d$ as the number of data packets successfully transmitted in one time slot and NTC($d$) is the number of transmission attempts required to achieve this. For the theoretical bounds, the capacity-cost function $C(\mathcal{P})$, given by Theorem 1 and Corollary 2, is plotted as a function of the power constraint $\mathcal{P}$. Here $\mathcal{P}$ is equivalent to the number of packet transmission attempts made in a time slot. The capacity-cost function $C(\mathcal{P})$ is divided by the packet length $N$ to represent the capacity in terms of data packets and offer a direct comparison to the NTC.

*1) Comparison of DP-Schedulers for FSMCs:* Figures 5 and 6 show the performance comparison of various DP-schedulers for 2- and 3-state Markov channels. The transition probabilities for the 2-state Markov channel (2SMC) are $p_{21} = 0.1$ and $p_{12} = 0.2$. The loss probability in each state is $\epsilon_1 = 0.2$ and $\epsilon_2 = 0.8$. For the 3-state Markov channel (3SMC), the transition probabilities are $p_{12} = 0.025$, $p_{13} = 0.075$, $p_{21} = 0.075$, $p_{23} = 0.05$, $p_{31} = 0.05$ and $p_{32} = 0.05$. The loss probabilities are $\epsilon_1 = 0.2$, $\epsilon_2 = 0.85$ and $\epsilon_3 = 0.95$. The time horizon $T$ for the simulations is 500 time slots and $10^5$ realizations of the channel are simulated. The packet error process is simulated by first generating a Markov process 500 time slots long and then setting the loss variable $z_k$ according to the given state $s_k$.

First, we plot the performance of DP-scheduler (solid red trace) and the corresponding bound (dashed red trace) for the non-causal perfect CSI case. For both cases (2SMC and 3SMC), the bound is tight up to the point when the deadline constraint requires a throughput of 0.2 packets/time slot. This can be due to the fact that for a lightly loaded system, the effect of queue state on decision making is minimal, and the scheduler considers CSI for decision making leading to a performance close to the theoretic bound. In all performance curves with feedback, for a 2SMC, a knee-point is observed when the NTC (or $\mathcal{P}$) is equal to 0.33 (for 3SMC, knee points at NTC = 0.37 and NTC = 0.6). As the number of packets to be transmitted is increased, the cost will increase at a certain rate since only the good states are being used initially. However after all the good states are exhausted, the scheduler will use the bad states for transmission, which increases cost at a faster rate as more packets will be dropped. For a throughput requirement of 0.25 packets/time slot in a 2SMC, causal perfect CSI scheduler requires 25% more transmission attempts than a non-causal perfect CSI scheduler (10% for a 3SMC).

We observe that the performance of the causal imperfect CSI scheduler or the ACK/NAK scheduler (solid black traces) is within 80% of the theoretical bound (dashed black traces) for the 2SMC (75% for the 3SMC). The bound is not very tight as the capacity calculations assume knowledge of the complete ACK/NAK sequence, whereas the ACK/NAK scheduler has access to the incomplete sequence (since ACK/NAK is unavailable for the time slots when no transmission is attempted). In both the cases, the ACK/NAK scheduler, at worst, requires 50% more transmission attempts than the non-causal perfect CSI based scheduler for achieving a throughput of 0.25 packets/time slot. The green trace represents a blind transmission scheme. In this case, the scheduler does not have any channel feedback, instead it just keeps transmitting till the queue is emptied out. This can be the lower bound on performance of any feedback scheme. The performance of this scheduler has a constant slope, since the scheduler does not adapt the transmission policy to the channel state. For a 2SMC, ACK/NAK scheduler outperforms the blind transmission scheme by up to 25% for 0.1 packets/time slot throughput (50% for a 3SMC) and 15% for 0.25 packets/time slot throughput (20% for a 3SMC).

*2) Look-up Table Size:* Figure 7 compares the effect of look-up table size on the performance of the ACK/NAK scheduler. The parameters for the 2SMC are: $p_{12} = 0.18$, $p_{21} = 0.09$, $\epsilon_1 = 0.23$ and $\epsilon_2 = 0.78$. The 32 MB table contains control action for $|\mathcal{A}| = 256$ channel realizations and 20 levels of $\boldsymbol{w_k}$. The parameters for the 1 MB table are $|\mathcal{A}| = 16$ and 10 levels of $\boldsymbol{w_k}$. Both tables store control action for upto 100 packets transmitted over a duration of upto 500 time slots. We observe that the ACK/NAK schedulers using a look-up table perform close to the exact solution. For instance, the 32 MB table requires 7% extra transmission attempts (11.5% for 1 MB table) to achieve throughput of 0.1 packets/time slot. Note that we have used uniform quantization for this example, however quantization methods that perform better might also exist.

*3) Time Horizon:* Figure 8 compares the performance of the ACK/NAK scheduler with different time horizons $T$. The parameters for the 2SMC are: $p_{12} = 0.2$, $p_{21} = 0.1$, $\epsilon_1 = 0.1$ and $\epsilon_2 = 0.9$. We observe that with increasing $T$, the performance of the ACK/NAK scheduler improves. A scheduler with $T = 100$ requires, at most, 10% more transmission attempts than a scheduler with $T = 800$. The improvement in performance can be due to the scheduler having greater flexibility in scheduling transmissions with larger $T$. We also observe that the performance of the ACK/NAK schedulers (for all $T$) is within 75% of the theoretical bound.

*4) Terminal Cost:* In Figure 9, we compare the effect of terminal cost $C$ on the performance of the ACK/NAK scheduler for a 2SMC. The channel parameters are: $p_{12} = 0.2$, $p_{21} = 0.1$, $\epsilon_1 = 0.2$ and $\epsilon_2 = 0.8$. For $C = \{10, 50, 200\}$, we observe that there is not much of a difference in the performance of the ACK/NAK schedulers formulated using these terminal costs. Other channel realizations also exhibit a similar behavior. This leads to the observation that the choice of $C$ does not impact the performance of the scheduler, as long as $C > 10$.

## B. Power Penalty

*1) Maximum Power Penalty:* We define the maximum power penalty for a feedback scheme using ACK/NAK information as $A_{\mathrm{dB}} \triangleq \max_R 10 \log_{10} \left( \frac{\text{Cost-Capacity for rate } R \text{ using ACK/NAK feedback}}{\text{Cost-Capacity for rate } R \text{ using causal CSI feedback}} \right)$. The Cost-Capacity function is the inverse of the Capacity-Cost function. Maximum power penalty quantifies the theoretical worst-case extra power required by an ACK/NAK based scheme to achieve the same throughput achieved by a causal perfect CSI based scheme. We define a similar quantity for the blind transmission scheme. Figure 10 compares the maximum power penalties incurred by ACK/NAK and blind schemes for 2SMCs as a function of channel memory $\mu \triangleq 1 - p_{12} - p_{21}$. We plot this quantity for different values of loss probabilities $\{\epsilon_1, \epsilon_2\}$, while keeping the steady state probabilities $\{\pi(1), \pi(2)\}$ constant. The ACK/NAK based schemes do not incur a maximum power penalty of more than 1 dB. On the other hand, in case of no feedback (blind) the maximum power penalty can be as much as 4.5 dB. An interesting observation is that when the channel conditions are "good" ($\epsilon_1 = 0.01$, $\epsilon_2 = 0.1$) the advantage of using an ACK/NAK based scheme is negligible. In this case, the difference in $A_{\mathrm{dB}}$ for the two cases is less than 0.1 dB. This can be explained by noting that under good channel conditions, there will be negligible packet drops and consequently, the value of estimating the channel is minimal.

*2) Actual Power Penalty:* Next, we find the actual power penalty incurred by an ACK/NAK scheduler, $L_{\mathrm{dB}} \triangleq 10 \log_{10} \left( \frac{\text{attempts made by ACK-NAK scheduler}}{\text{attempts made by causal CSI scheduler}} \right)$, for a given throughput. We define a similar quantity for the blind transmission scheme. Figure 11 shows the power penalty incurred by the ACK/NAK based and blind schedulers for 2SMCs as a function of channel memory $\mu$. We plot this quantity for different values of packet loss probabilities, $\{\epsilon_1, \epsilon_2\}$ keeping the steady state probabilities $\{\pi(1), \pi(2)\}$ constant. For the range of channel parameters considered in our simulations and a throughput requirement of 0.2 packets/time slot, ACK/NAK scheduler does not incur a power penalty of more than 1.5 dB. On the other hand, a blind transmission scheme takes a power penalty of up to 4.5 dB (for $\epsilon_1 = 0$ and $\epsilon_2 = 1$).

In Figure 12, we consider different steady state probabilities $\{\pi(1), \pi(2)\}$, while keeping the loss probabilities constant ($\epsilon_1 = 0.1, \epsilon_2 = 0.9$). The throughput requirement is set at 0.1 packets/time slot. In the worst case, the ACK/NAK scheduler incurs a power penalty of 2.5 dB for $\pi_1 = 0.2$ and $\pi_2 = 0.8$ whereas the blind scheme incurs a power penalty of almost 4.5 dB. For $\mu = 0.9$, in all the cases, the performance of the ACK/NAK scheduler is 2 dB better than the blind scheduler. Since $\mu$ is an indicator of the burstiness (as bursty channels have high values of $\mu$), the improvement provided by our scheme will be more pronounced in bursty channels.

## VI. Conclusions

We introduced a channel aware transmission scheduler to optimize energy efficiency in sensor networks. Our transmission scheduler meets the deadline constraint for all packets waiting in the transmission queue while optimizing the total energy spent on transmission. To do this, we used finite-horizon DP and utilized different types of CSI in the optimization process. We also derived the capacity-cost function of a FSMC with different types of CSI at the controller. The capacity-cost function can be used as a bound for the performance of the transmission schedulers. We showed that the performance achieved by a DP-based scheduler is close to the fundamental limits for a wide range of channel parameters. Indeed, for some cases, the ACK/NAK scheduler achieves performance that is within 80% of the theoretical bound. Comparisons between the performance of different grades of CSI show that the difference between performance of causal perfect CSI and ACK/NAK schemes for a wide range of channel parameters is not significant. On the other hand, blind transmission scheme in which the ACK/NAK feedback is not exploited to estimate the channel state incurs a significant power penalty. In addition, the maximum power penalty incurred by ACK/NAK feedback based schemes provide insights into the fundamental value of using ACK/NAK information while scheduling transmissions. For instance, in one case, the maximum power penalty of not using any kind of feedback is 4.5 dB. The use of ACK/NAK information, which is available to the transmitter in most wireless link layer protocols without extra effort, can reduce this power penalty to less than 1 dB.

## References

[1] M. Zorzi and R. R. Rao, "Error Control and Energy Consumption in Communications for Nomadic Computing," *IEEE Trans. Comput.*, vol. 46, no. 3, pp. 279–289, Mar. 1997.

[2] C.-F. Chiasserini and M. Meo, "Impact of ARQ Protocols on QoS in 3GPP Systems," *IEEE Trans. Veh. Technol.*, vol. 52, no. 1, pp. 205–215, Jan. 2003.

[3] L. A. Johnston and V. Krishnamurthy, "Opportunistic File Transfer over a Fading Channel: A POMDP Search Theory Formulation with Optimal Threshold Policies," *IEEE Trans. Wireless Commun.*, vol. 5, no. 2, pp. 394–405, February 2006.

[4] E. Uysal-Biyikoglu, B. Prabhakar, and A. E. Gamal, "Energy-Efficient Packet Transmission Over a Wireless Link," *IEEE/ACM Trans. Networking*, vol. 10, no. 4, pp. 487–499, August 2002.

[5] M. Zafer and E. Modiano, "Delay-Constrained Energy Efficient Data Transmission over a Wireless Fading Channel," in *Proc. Information Theory and Applications Workshop '07*, San Diego, USA, Jan. 2007.

[6] M. A. Haleem and R. Chandramouli, "Adaptive Downlink Scheduling and Rate Selection: A Cross-layer Design," *IEEE J. Select. Areas Commun.*, vol. 23, no. 6, pp. 1287– 1297, June 2005.

[7] C. K. Ho and J. Oostveen, "Rate Adaptation in Time Varying Channels using Acknowledgement Feedback," in *Proc. of IEEE VTC'06-Spring*, May 2006, pp. 1683–1687.

[8] A. K. Karmokar, D. V. Djonin, and V. K. Bhargava, "POMDP-Based Coding Rate Adaptation for Type-I Hybrid ARQ Systems over Fading Channels with Memory," *IEEE Trans. Wireless Commun.*, vol. 5, no. 12, pp. 3512–3523, Dec. 2006.

[9] D. V. Djonin, A. K. Karmokar, and V. K. Bhargava, "Joint Rate and Power Adaptation for Type-I Hybrid ARQ Systems Over Correlated Fading Channels Under Different Buffer-Cost Constraints," *IEEE Trans. Veh. Technol.*, vol. 57, no. 1, pp. 421–435, Jan. 2008.

[10] A. K. Karmokar, D. V. Djonin, and V. K. Bhargava, "Cross-Layer Rate and Power Adaptation Strategies for IR-HARQ Systems over Fading Channels with Memory: A SMDP-Based Approach," *IEEE Trans. Commun.*, vol. 56, no. 8, pp. 1352–1365, August 2008.

[11] A. J. Goldsmith and P. P. Varaiya, "Capacity of Fading Channels with Channel Side Information," *IEEE Trans. Inform. Theory*, vol. 43, no. 6, pp. 1986–1992, November 1997.

[12] T. M. Cover and J. A. Thomas, *Elements of Information Theory*, 1st ed. John Wiley and Sons, 1999.

[13] H. Viswanathan, "Capacity of Markov Channels with Receiver CSI and Delayed Feedback," *IEEE Trans. Inform. Theory*, vol. 45, no. 2, pp. 761–771, March 1999.

[14] S. Yüksel and S. Tatikonda, "Capacity of Markov Channels with Partial State Feedback," in *Proc. IEEE ISIT '07*, Nice, France, June 2007, pp. 1861–1865.

[15] S. Verdú, "On Channel Capacity per Unit Cost," *IEEE Trans. Inform. Theory*, vol. 36, no. 5, pp. 1019–1030, Sept. 1990.

[16] E. N. Gilbert, "Capacity of a Burst-Noise Channel," *Bell Syst. Tech. J.*, vol. 39, pp. 1253–1266, Sept. 1960.

[17] P. Sadeghi, R. A. Kennedy, P. B. Rapajic, and R. Shams, "Finite-State Markov Modeling of Fading Channels - A Survey of Principles and Applications," *IEEE Signal Process. Mag.*, vol. 25, no. 5, pp. 57–80, September 2008.

[18] L. R. Rabiner, "A Tutorial on Hidden Markov Models and Selected Applications in Speech Recognition," *Proc. IEEE*, vol. 77, no. 2, pp. 257–286, February 1989.

[19] D. P. Bertsekas, *Dynamic Programming and Optimal Control*, 3rd ed. Athena Scientific, 2005.

[20] R. J. McEliece, *The Theory of Information and Coding*, 2nd ed. Cambridge University Press, 2002.

[21] R. A. Berry and R. G. Gallager, "Communication Over Fading Channels With Delay Constraints," *IEEE Trans. Inform. Theory*, vol. 48, no. 5, pp. 1135–1149, May 2002.

APPENDIX

### A. Proof for Theorem 1

The capacity-cost function of a cost constrained memoryless stationary channel is defined as, $C(\mathcal{P}) \triangleq \sup_{\substack{X \\ E[b[X]] \leq \mathcal{P}}} I(X;Y)$. Where $b[X]$ is the cost of transmitting symbol $X$. The transmission of one bit of information requires $1/C(\mathcal{P})$ symbols at a cost of $\mathcal{P}/C(\mathcal{P})$. In order to find $C(\mathcal{P})$, first we find the capacity per unit cost, $C_{\text{Cost}}$, using the expression in [15],

$$C_{\text{cost}} = \sup_X \frac{I(X;Y)}{E[b(X)]} = \sup_x \frac{D(P_{Y|X=x}||P_{Y|X='v'})}{b(x)} = (1 - \epsilon_s)N. \qquad (9)$$

In our case, $b(x) = 1$ for $x = \{`1`, \ldots, `2^N`\}$ and $b(x) = 0$ for $x = `\mathcal{V}`$. We find that for an ECV, $D(P_{Y|X=`1`}||P_{Y|X=`\mathcal{V}`}) = \ldots = D(P_{Y|X=`2^N`}||P_{Y|X=`\mathcal{V}`}) = 1 - \epsilon_s$. Using Eq. (9) and incorporating the power constraint of $\mathcal{P}$ we can find the capacity-cost function of the ECV,

$$C(\mathcal{P}) = E[b(X)]C_{\text{cost}} = \mathcal{P}\left(N(1 - \epsilon_s)\right). \tag{10}$$

For an average power constraint $\mathcal{P}$ and unit feedback delay, $C_{\text{fb}}(\mathcal{P})$ is given by [13],

$$C_{\text{fb}}(\mathcal{P}) = \max_{P(X|\tilde{S})} I(X;Y|S,\tilde{S}), \text{ s.t. } \sum_{\tilde{s}} \pi(\tilde{s})\mathcal{P}(\tilde{s}) \leq \mathcal{P}, \ 0 \leq \mathcal{P}(\tilde{s}) \leq \mathcal{P}, \ \forall \tilde{s}. \tag{11}$$

Where $\tilde{s} \in \mathcal{S}$ represents the state information fed back ($\tilde{s} = s_{t-1}$) to the transmitter from the receiver and $s \in \mathcal{S}$ is state at the time of transmission. $\mathcal{P}(\tilde{s})$ is the average cost per symbol incurred while in state $\tilde{s}$. And we define,

$$I(X;Y|S,\tilde{S}) = \sum_{\tilde{s}\in\mathcal{S}} \pi(\tilde{s})I(X;Y|S,\tilde{S}=\tilde{s}) = \sum_{\tilde{s}\in\mathcal{S}} \pi(\tilde{s})\sum_{s\in\mathcal{S}} \mathrm{P}(\tilde{s},s)I(X;Y|S=s,\tilde{S}=\tilde{s}).$$

Using the capacity-cost function from Eq. (10) in conjunction[2] with the capacity expression (11), we have the capacity-cost function for FSMECs,

$$C_{\text{FSMEC}}(\mathcal{P}) = \sup_{\mathcal{P}(\tilde{s})} \sum_{\tilde{s}} \sum_{s} \pi(\tilde{s})\mathrm{P}(\tilde{s},s)(1-\epsilon_s)N\mathcal{P}(\tilde{s}), \text{ s.t. } \sum_{\tilde{s}} \pi(\tilde{s})\mathcal{P}(\tilde{s}) \leq \mathcal{P}, \ 0 \leq \mathcal{P}(\tilde{s}) \leq \mathcal{P}, \ \forall \tilde{s}.$$

### B. Proof for Corollary 2

We calculate the capacity of an energy-constrained finite state Markov erasure channel with causal imperfect CSI by extending the proof given by Yüksel and Tatikonda in [14]. The problem formulation in that work is very similar to ours. The main difference comes from the fact that we consider a causal probabilistic quantizer (i.e., ACK/NAK) for state feedback. In order to incorporate this feature, we first make some modifications to converse of the channel theorem.

If we consider a coding scheme such that $e \to 0$, Fano's inequality gives,

$$H(W|Y_1^T, S_1^T, Z_1^T) \leq h(p_e) + p_e \log_2(M), \tag{12}$$

where $p_e$ gives the probability of error. Also,

$$H(W|Y_1^T, S_1^T, Z_1^T) = H(W) - I(W;Y_1^T, S_1^T, Z_1^T) = \log_2(M) - I(W;Y_1^T, S_1^T, Z_1^T). \tag{13}$$

---

[2]It might appear that the capacity-cost calculations which have been made for a memoryless channel are not applicable for Markov channels (which are not memoryless). However [13] derives capacity results which use the mutual information rate conditioned on the present state of the channel, which makes the channel memoryless for Markov channels.

Combining Eqs. (12) and (13) we have,

$$\frac{\log_2(M)}{T} \leq \frac{I(W; Y_1^T, S_1^T, Z_1^T) + h(p_e)}{T(1 - p_e)}.$$

As $T \to \infty$, $p_e \to 0$ and we can write,

$$\limsup_{N \to \infty} \frac{\log_2(M)}{T} \leq \limsup_{N \to \infty} \frac{1}{T} I(W; Y_1^T, S_1^T, Z_1^T). \tag{14}$$

We focus our attention on the right side of Eq. (14),

$$\frac{1}{T} I(W; Y_1^T, S_1^T, Z_1^T)$$

$$\overset{(a)}{=} \frac{1}{T} \sum_{k=1}^{T} I(W, Z_1^{k-1}; Y_k, S_k, Z_k | Y_1^{k-1}, S_1^{k-1}, Z_1^{k-1}) - \underbrace{I(Z_1^{k-1}; Y_k, S_k, Z_k | Y_1^{k-1}, S_1^{k-1}, Z_1^{k-1}, W)}_{=0}$$

$$\overset{(b)}{=} \frac{1}{T} \sum_{k=1}^{T} H(Y_k, S_k, Z_k | Y_1^{k-1}, S_1^{k-1}, Z_1^{k-1}) - H(Y_k, S_k, Z_k | Y_1^{k-1}, S_1^{k-1}, Z_1^{k-1}, W, X_k)$$

$$\overset{(c)}{=} \frac{1}{T} \sum_{k=1}^{T} H(S_k | Y_1^{k-1}, S_1^{k-1}, Z_1^{k-1}) + H(Y_k | Y_1^{k-1}, S_1^{k-1}, Z_1^{k-1}, S_k)$$

$$+ H(Z_k | Y_1^{k-1}, S_1^{k-1}, Z_1^{k-1}, S_k, Y_k) - H(S_k | S_1^{k-1}, Z_1^{k-1}, W, X_k)$$

$$- H(Y_k | S_1^{k-1}, Z_1^{k-1}, W, X_k, S_k) - H(Z_k | S_1^{k-1}, Z_1^{k-1}, W, X_k, S_k, Y_k)$$

$$= \frac{1}{T} \sum_{k=1}^{T} H(Y_k | Y_1^{k-1}, S_1^{k-1}, Z_1^{k-1}, S_k) + \underbrace{H(S_k | S_{k-1})}_{(d)} + H(Z_k | S_k) - H(S_k | S_{k-1})$$

$$- H(Y_k | S_k, X_k) - \underbrace{H(Z_k | X_k, Y_k)}_{(e)}$$

$$= \frac{1}{T} \sum_{k=1}^{T} \underbrace{H(Y_k | Y_1^{k-1}, S_1^{k-1}, Z_1^{k-1}, S_k) - H(Y_k | S_k, U_k)}_{\text{I}} + \underbrace{H(Z_k | S_k)}_{\text{II}}$$

Here, $(a)$ and $(c)$ follow from the chain rule of Mutual Information and entropy respectively, and $(b)$ is due the fact that the channel input $X_k$ is a function of the channel observations $Z_1^{k-1}$ and the message $W$. The above expression can be divided into two parts I and II. Part I is the same as that given by [14] and can be written as a cost that is a function of the coding policy $p(x_k | z_1^{k-1})$. This cost expression is used to recast the maximization problem (14) as a Markov Decision Process. Consequently, the argument for the converse just follows the original proof given in [14]. Part II of the expression is a constant that does not depend on the coding policy and can be ignored. The direct theorem can be used without modification to complete the proof.

$u_k$

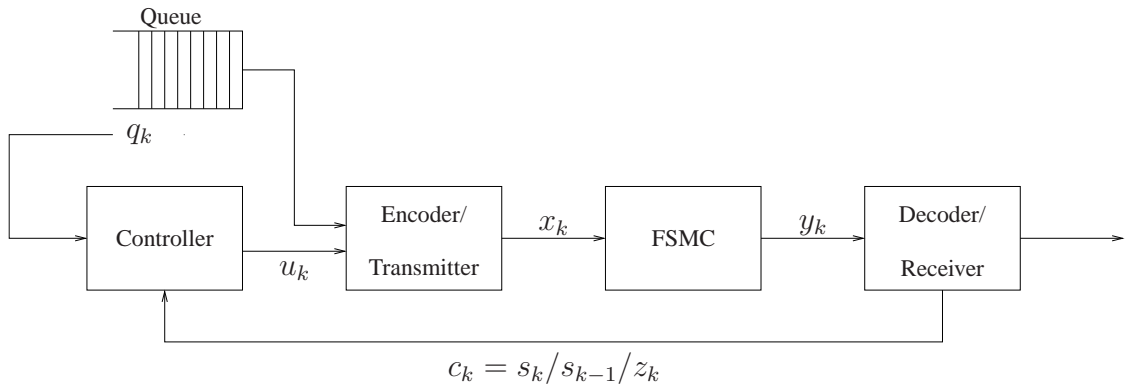

Fig. 1. Block diagram showing the system model considered in this work.

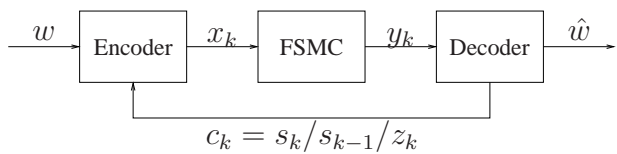

Fig. 2. Block diagram showing the system model used for the information theoretic capacity calculations for the FSMC with CSI at the receiver and feedback.

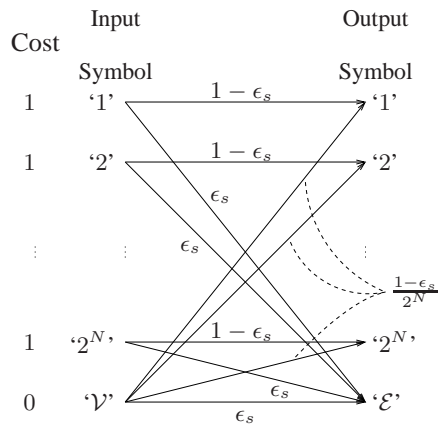

Fig. 3. The Erasure Channel with Vacations (ECV). '$\mathcal{V}$' is a zero-cost, zero-information symbol. {'1','2',…,'$2^N$'} incur a unit cost on transmission. The transition labels represent the appropriate probabilities.

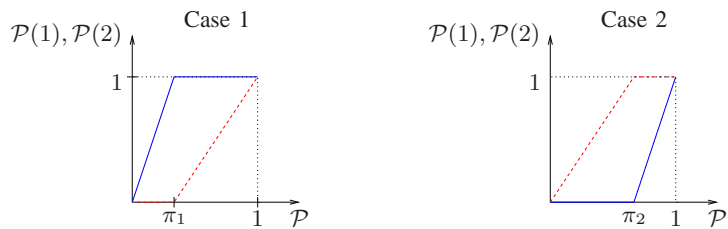

Fig. 4. The power allocation policy as a function of the power constraint $\mathcal{P}$ for Case 1 of the 2SMEC with perfect causal CSI. The solid (dashed) trace corresponds to the power allocation for $\tilde{s} = 1(2)$.

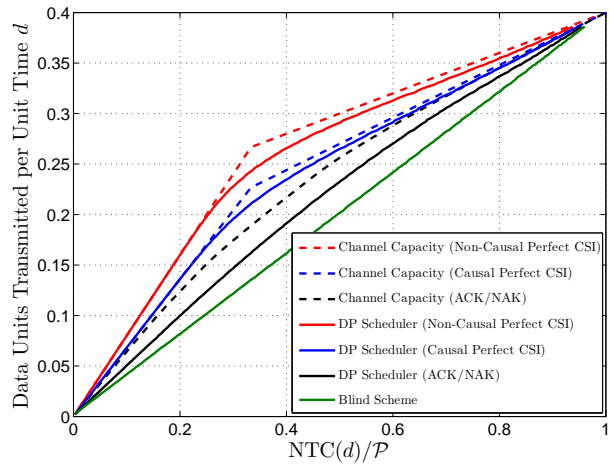

Fig. 5. Comparison of the energy efficiency performance of various schemes with their theoretical bounds for 2-state Markov channel. Here $\epsilon_1 = 0.2$, $\epsilon_2 = 0.8$, $p_{21} = 0.1$ and $p_{12} = 0.2$.

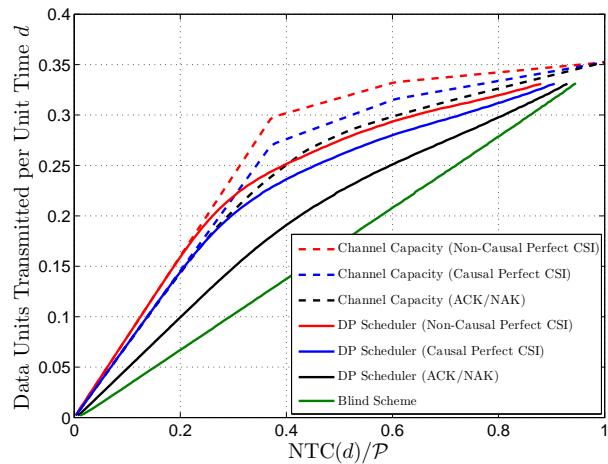

Fig. 6. Comparison of the energy efficiency performance of various schemes with their theoretical bounds for 3-state Markov channel. Here $p_{12} = 0.025$, $p_{13} = 0.075$, $p_{21} = 0.075$, $p_{23} = 0.05$, $p_{31} = 0.05$ and $p_{32} = 0.05$. The loss probabilities are $\epsilon_1 = 0.2$, $\epsilon_2 = 0.85$ and $\epsilon_3 = 0.95$.

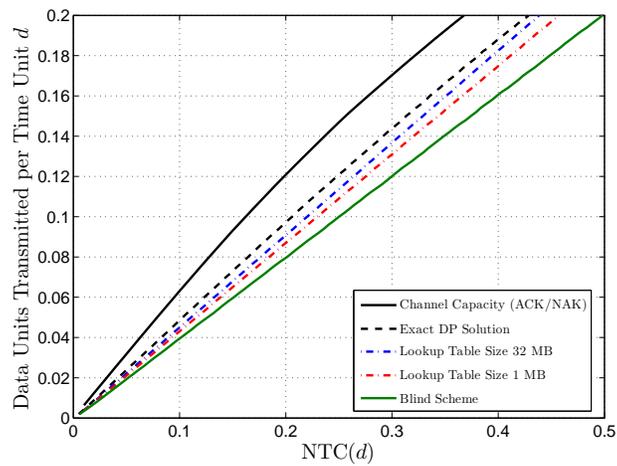

Fig. 7. Comparison of the performance of ACK/NAK schedulers with different table sizes. Here $\epsilon_1 = 0.23$, $\epsilon_2 = 0.78$, $p_{21} = 0.09$ and $p_{12} = 0.18$.

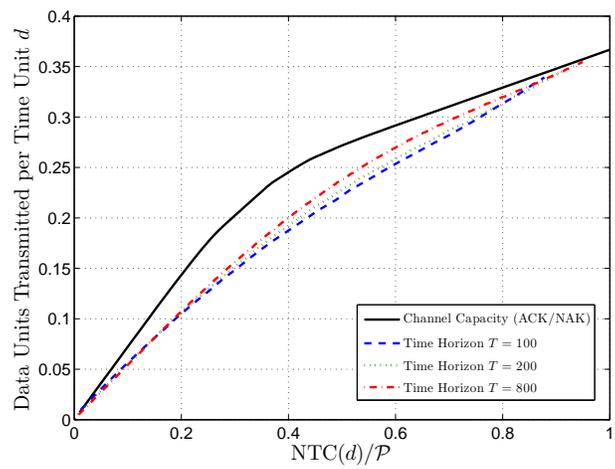

Fig. 8. Comparison of the performance of ACK/NAK schedulers with different time horizons $T$. Here $\epsilon_1 = 0.1$, $\epsilon_2 = 0.9$, $p_{21} = 0.1$ and $p_{12} = 0.2$.

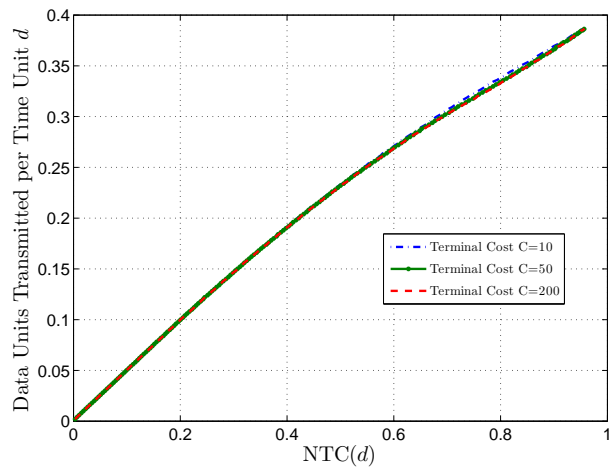

Fig. 9.  Comparison of the performance of ACK/NAK schedulers formulated with different terminal costs $C$. Here $\epsilon_1 = 0.2$, $\epsilon_2 = 0.8$, $p_{21} = 0.1$ and $p_{12} = 0.2$.

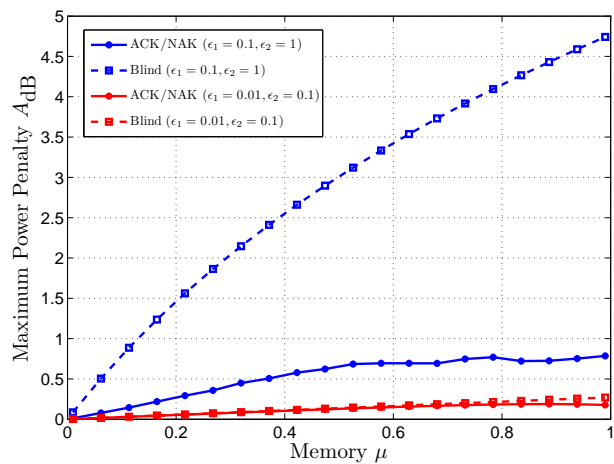

Fig. 10.  The maximum power penalty $A_{\mathrm{dB}}$ as a function of the channel memory $\mu$ for different values of packet loss probabilities $\{\epsilon_1, \epsilon_2\}$ with $\pi(1) = 1/3$ and $\pi(2) = 2/3$.

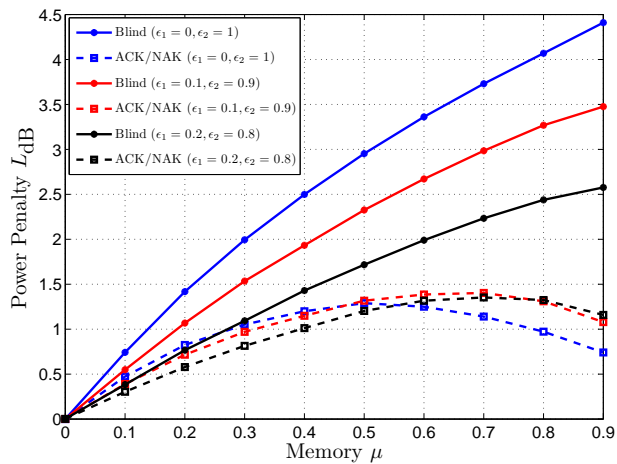

Fig. 11.   The power penalty $L_{\text{dB}}$ as a function of the channel memory $\mu$ for different values of packet loss probabilities $\{\epsilon_1, \epsilon_2\}$ with $\pi(1) = 1/3$ and $\pi(2) = 2/3$. The scheduler attempts to transmit 100 packets over 500 time slots for all cases.

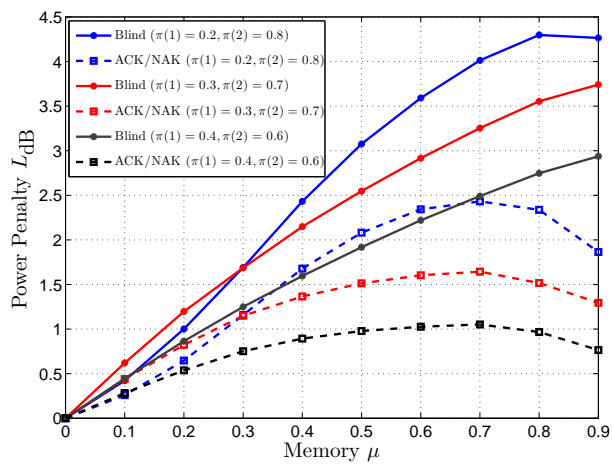

Fig. 12.   The power penalty $L_{\text{dB}}$ as a function of the channel memory $\mu$ for different values of steady state probabilities $\{\pi(1), \pi(2)\}$ with $\epsilon_1 = 0.1$ and $\epsilon_2 = 0.9$. The scheduler attempts to transmit 50 packets over 500 time slots for all cases.